\documentclass[showpacs,superscriptaddress,floatfix,amsmath,amssymb,twocolumn,pra]{revtex4-1}
\usepackage{bm}
\usepackage{graphicx}
\usepackage{color}
\usepackage[normalem]{ulem}
\usepackage{epstopdf}

\def\sgn{{\rm sgn}}

\newcommand{\fp}{f_{\rm p}}
\newcommand{\mup}{\mu_{\rm p}}
\newcommand{\lap}{\lambda_{\rm p}}

\def\dag{\dagger}

\begin{document}
\title {Critical fluctuations and the rates of interstate switching near excitation threshold of a quantum parametric oscillator 
}
\author{Z. R. Lin}
\affiliation{RIKEN Center for Emergent Matter Science (CEMS), Wako, Saitama 351-0198, Japan}
\author{Y. Nakamura}
\affiliation{RIKEN Center for Emergent Matter Science (CEMS), Wako, Saitama 351-0198, Japan}
\affiliation{Research Center for Advanced Science and Technology (RCAST), The University of Tokyo, Meguro-ku, Tokyo 153-8904, Japan}
\author{M. I. Dykman}
\affiliation{Department of Physics and Astronomy, Michigan State University, East Lansing, MI 48824, USA}

\date{\today}

\begin{abstract}

We study the dynamics of a nonlinear oscillator near the critical point where period-two vibrations are first excited with the increasing amplitude of parametric driving. Above the threshold, quantum fluctuations induce transitions between the period-two states over the quasienergy barrier. We find the effective quantum activation energies for such transitions and their scaling with the difference of the driving amplitude from its critical value. We also find the scaling of the fluctuation correlation time with the quantum noise parameters in the critical region near the threshold.

\end{abstract}


\maketitle

\section{Introduction}

Parametrically driven oscillators are of fundamental interest as a platform for studying quantum phenomena away from thermal equilibrium in well-characterized systems. Examples of such phenomena include noise squeezing, low-noise quantum amplification, photon generation, quantum measurements, and coherent state preparation in a dissipative environment \cite{Yurke1988,Castellanos-Beltran2008,Yamamoto2008a,Wilson2010,Krantz2013,Lin2014,Eichler2014,Leghtas2015}. Classically, parametrically excited vibrational states are states with a broken discrete time-translation symmetry, as their period is twice the modulation period \cite{LL_Mechanics2004}. Excitation of period two vibrations with the increasing driving strength is somewhat reminiscent of the mean-field ferromagnetic phase transition with lowering temperature.

A distinctive feature of the parametric oscillator is the possibility to have  a stable state with unbroken time-translation symmetry along with the stable period-two vibrational states (more precisely, such states are called asymptotically stable). In the symmetric state, vibrations are either not excited or there may be small-amplitude vibrations with the same period as the driving. In the parameter space there is a critical point where this state and the period-two states merge.

In this paper we study quantum fluctuations near the critical point. They are strong and slow. We find the long-time behavior of their correlation functions and the scaling of the correlation time with $\hbar$ and temperature. As the system moves deeper into the range of coexisting period-two states, in the neglect of fluctuations it would localize in one of them or in the symmetric state, if it is also stable. A major manifestation of quantum fluctuations is switching between the states. We identify the switching mechanism and find the switching rates. We study how the rates scale with the driving amplitude and how there occurs a crossover to the previously explored scaling far from the critical point. 

We focus on quantum effects for weakly damped oscillators, where the decay rate $\Gamma$ is small compared to the oscillator eigenfrequency $\omega_0$. This condition is usually met in the experiments with superconducting microcavities and with Josephson-junction based systems used in quantum information. For weakly damped oscillators, nonlinearity becomes substantial already for comparatively small vibration amplitudes, once the nonlinearity-induced shift of the vibration frequency becomes comparable with $\Gamma$. It is the nonlinearity that leads to multistability of forced vibrations. Even though the nonlinear effects are strong, they occur in the range of comparatively small vibration amplitudes, and classically, the oscillator motion is almost sinusoidal vibrations with slowly varying amplitude and phase.

\begin{figure}[h]
\includegraphics[width=6cm]{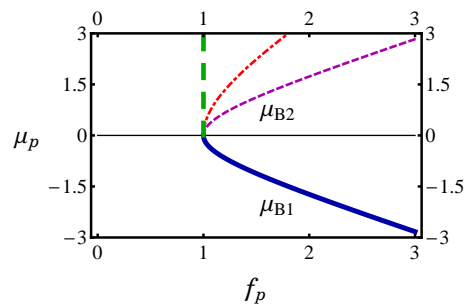}
\caption{The relation between the parameters of the driving field where the number of stable vibrational states changes. Parameters $\mup$ and $\fp$ characterize the scaled detuning of the field frequency from resonance and the scaled field amplitude, respectively, see Eq.~(\ref{eq:g_parametric}). For $\fp<1$ or $\mup<\mu_{B1}$ the oscillator has no period-two vibrational states. In the region $\fp>1$ and $\mup>\mu_{B1}$ there are two stable period-two states. For $\mup>\mu_{B2}$ the state with no period-two vibrations is stable, too. On the long-dashed line the stable and unstable period-two states coalesce. The point $\mup=0,\fp=1$ where the three bifurcation lines merge is the critical point. The dot-dashed line shows where the three stable states are equally occupied in the stationary regime.}
\label{fig:bifurcation}
\end{figure}

Classically, the dynamics of a driven underdamped oscillator is determined by two parameters, the scaled frequency and amplitude of the driving \cite{LL_Mechanics2004}. The analysis of the dynamics is simplified near bifurcation parameter values, where the number of the stable vibrational states changes. In this region one of the motions becomes slow, an analog of the ``soft mode" \cite{Guckenheimer1987}. The bifurcation relation between the oscillator parameters in the case of close-to-resonant parametric driving is shown in Fig.~\ref{fig:bifurcation}.

The possibility to describe the dynamics by a single dynamical variable also significantly simplifies quantum analysis \cite{Dykman2007,*Dykman2012}. Such variable commutes with itself at different times, and therefore its dynamics is essentially classical. However, the fluctuations are quantum, because the noise that causes them has quantum origin and its intensity is $\propto \hbar$ for low temperature.  As we show, this picture applies near the critical point. We note that, in terms of the Floquet states of the modulated oscillator, this case corresponds to the distance between the quasienergy levels being much smaller than their width due to dissipation.

\section{The model}

We consider a parametrically driven oscillator with the Duffing (Kerr) nonlinearity. 
The Hamiltonian of the oscillator reads 
\begin{align}
\label{eq:H_0(t)}
H_0=\frac{1}{2}p^2+\frac{1}{2}\omega_0^2q^2 +\frac{1}{4}\gamma q^4 + \frac{1}{2}q^2F\cos \omega_Ft,
\end{align}
where $q$ and $p$ are the oscillator coordinate and momentum, the mass is set equal to one, and
$\gamma$ is the anharmonicity parameter. Parameter $F$ gives the amplitude of the driving, whereas $\omega_F$ is the driving frequency. 

We assume the driving to be resonant and not too strong,
\begin{equation}
\label{eq:delta_omega}
|\delta\omega|\ll \omega_0, \qquad \delta\omega=\frac{1}{2}\omega_F-\omega_0; \qquad |\gamma|\langle q^2
\rangle\ll\omega_0^2.
\end{equation}

It is convenient to change to the rotating frame using the standard canonical transformation $U(t)=\exp\left(-ia^\dagger a\,\omega_Ft/2\right)$, where $a^\dagger$ and $a$ are the raising and lowering operators of the oscillator. We introduce slowly varying in time dimensionless coordinate $Q$ and momentum $P$, using as a scaling factor the characteristic amplitude of forced vibrations $C_{\rm p}=|2F_{\rm c}/3\gamma|^{1/2}$ [parameter $F_{\rm c}$ is defined below in Eq.~(\ref{eq:critical_field})],
\begin{eqnarray}
\label{eq:canonical_transform}
U^{\dag}(t)q U(t) &=& C_{\rm p}(Q\cos\varphi_{}+P\sin\varphi_{}), \nonumber\\
U^{\dag}(t)p U(t) &=& -\frac{1}{2}C_{\rm p}\omega_F(Q\sin\varphi_{} - P\cos\varphi_{})
\end{eqnarray}
with $\varphi_{}=\frac{1}{2}\omega_Ft+\frac{1}{4}\pi$. The commutation relation between $P$ and $Q$ is
\begin{equation}
\label{eq:lambda}
[P,Q]=-i\lambda_{\rm p},\qquad \lambda_{\rm p}=3|\gamma|\hbar/(\omega_F F_{\rm c}) .
\end{equation}
Parameter $\lambda_{\rm p}\propto \hbar$ plays the role of the Planck constant in the quantum dynamics in the rotating frame. It is determined by the oscillator nonlinearity, $\lambda_{\rm p}\propto \gamma$, 
For characteristic $|Q|,|P|\lesssim 1$, where $\langle q^2\rangle \lesssim C_{\rm p}^2$, the last inequality in Eq.~(\ref{eq:delta_omega}) implies  $F\ll \omega_0^2$. To simplify the analysis near the critical point,  in Eqs.~(\ref{eq:canonical_transform}) and (\ref{eq:lambda}) we choose the variables in the form that slightly differs from that in Ref.~\onlinecite{Dykman2007},

In the range (\ref{eq:delta_omega}) the oscillator dynamics can be analyzed in the rotating wave approximation (RWA). The Hamiltonian in the rotating frame is
$U^{\dag}H_0 U-i\hbar U^{\dag}\dot{U} =
(F_{\rm c}^2/6\gamma)\hat{g}_{\rm p}$. Operator $\hat g_{\rm p}=g_{\rm p}(Q,P)$ is independent of time and has the form
\begin{align}
\label{eq:g_parametric}
&g_{\rm p}(Q,P) =
\frac{1}{4}\left(P^2+Q^2\right)^2 - \frac{1}{2}\mu_{\rm p}(P^2+Q^2)\nonumber\\
&\qquad +\frac{1}{2}f_{\rm p}\sgn\gamma(QP+PQ), \nonumber\\
&\mu_{\rm p}=\frac{\omega_F(\omega_F-2\omega_0)}{F_{\rm c}}\sgn\gamma,\qquad
f_{\rm p}=F/F_{\rm c}.
\end{align}
The eigenvalues of $\hat{g}_{\rm p}$ (multiplied by $F_{\rm c}^2/6\gamma$) give the oscillator  quasienergies,

In Eq.~(\ref{eq:g_parametric}), parameter $f_{\rm p}$ is the scaled driving amplitude $F$; in what follows we assume $f_{\rm p}>0$. Parameter $\mu_{\rm p}$ gives the detuning of the driving frequency from twice the oscillator eigenfrequency. The scaling factor in both $f_{\rm p}$ and $\mu_{\rm p}$ is the critical amplitude $F_{\rm c}$ needed for parametric excitation of the oscillator. 

Relaxation of the oscillator results from the coupling to a thermal bath. The coupling leads to oscillator decay due to scattering by the bath excitations. For a weak coupling, one of the most important scattering mechanisms is scattering with energy transfer $\approx \hbar\omega_0$ in an elementary event \cite{Einstein1910b,Weisskopf1930,Schwinger1961}. It comes from the coupling, which is linear in the oscillator coordinate and/or momentum. In a phenomenological description of the oscillator dynamics such scattering  corresponds to a friction force $-2\Gamma\dot q$. The friction coefficient $\Gamma$ is simply expressed in terms of the appropriate correlator of the thermal bath variables. 

It is important for what follows that relaxation comes along with a quantum noise. The equations of motion for slow variables $Q,P$ in the rotating frame in dimensionless time $\tau = \Gamma t$ read
\begin{align}
\label{eq:q_Langevin}
&\dot Q=-i\lambda_{\rm p}^{-1}[Q,\hat g_{\rm p}]\sgn\gamma  - Q+ \hat f_Q(\tau),\nonumber\\
&\dot P=-i\lambda_{\rm p}^{-1}[P,\hat g_{\rm p}]\sgn\gamma  - P + \hat f_P(\tau).
\end{align}
Here, $\dot A \equiv dA/d\tau$. We have set
\begin{equation}
\label{eq:critical_field}
F_{\rm c}=2\Gamma\omega_F,
\end{equation}
in which case, indeed, $F_{\rm c}$ is the threshold value of the driving amplitude, as will be seen below. 

In Eq.~(\ref{eq:q_Langevin}), $\hat f_{Q,P}$ are quantum noise operators. The noise is $\delta$-correlated in slow time, cf. \cite{Dykman2012},
\begin{align}
\label{eq:noise_correlators}
&\langle \hat f_Q(\tau)\hat f_Q(\tau')\rangle_{\rm b}=\langle \hat f_P(\tau)\hat f_P(\tau')\rangle_{\rm b}= 2{\cal D}
\delta(\tau-\tau'),\nonumber\\
&\langle [\hat f_Q(\tau),\hat f_P(\tau')]\rangle_{\rm b}=2i\lambda_{\rm p}\delta(\tau-\tau'),
\nonumber\\
&{\cal D}=\lambda _{\rm p}(\bar n+1/2), \qquad  \bar n = [\exp(\hbar\omega_0/k_BT)-1]^{-1}.
\end{align}
Here, $\bar n$ is the oscillator Planck number and $\langle \cdot\rangle_{\rm b}$ means averaging over the bath state. Parameter ${\cal D}$ plays the role of the effective temperature of the quantum noise, ${\cal D}\propto \hbar$ for $\hbar\omega_0\gg k_BT$. The commutation relation in Eq.~(\ref{eq:noise_correlators}) guarantees that the commutator $[Q,P]=i\lambda_{\rm p}$ does not change in time. The noise correlators are understood in the Stratonovich sense \cite{vanKampen_book}; in particular,  $\langle[\hat f_Q(\tau),P(\tau)]\rangle_{\rm b} = \langle[Q(\tau),\hat f_P(\tau)]\rangle_{\rm b}=i\lambda_{\rm p}$.

\section{Slow quantum dynamics near the critical point}

The stable vibrational states of the oscillator are given by the stable stationary solutions of Eq.~(\ref{eq:q_Langevin}) without noise. The bifurcation values of the parameters $f_{\rm p}, \mu_{\rm p}$, where the number of the stable states changes, lie on the lines $\mup=\pm\mu_B(\fp)$ and $(\mup\geq 0,\fp=1)$ on the $(f_{\rm p}, \mu_{\rm p})$-plane,
\begin{equation}
\label{eq:bifurcation_line}
\mu_{B}= (\fp^2-1)^{1/2}.
\end{equation}
The bifurcation lines are shown in Fig.~\ref{fig:bifurcation}. For $\mu_{\rm p} < \mu _{B1} \equiv -\mu_{B}$ and for $\fp<1$ the only stable state of the oscillator is $Q=P=0$. The amplitude of period-two vibrations in this state is zero, and below we call it the zero-amplitude state. At $\mup=\mu_{B1}$ this state becomes unstable. For larger $\mup$ or $\fp$ the oscillator has  two stable states, which correspond to period-two vibrations in the lab frame with the reduced squared amplitude $Q^2+P^2= \mup +(\fp^2-1)^{1/2}$ and with the phases differing by $\pi$. For $\mup =\mu_{B2}$ the state $Q=P=0$ becomes stable and there emerge two unstable states, which correspond to unstable period-two vibrational states in the lab frame\cite{LL_Mechanics2004}. The values of the control parameter $\mu_{B1}$ and $\mu_{B2}$ correspond to the supercritical and subcritical pitchfork bifurcations \cite{Guckenheimer1987}. On the line $\fp=1,\mup>0$ the stable and unstable states, which correspond to period-two vibrations, coalesce.

At  the critical point $(\mup=0,\fp=1)$ all three bifurcation lines merge. Such merging is robust, it does not disappear if the model is slightly changed, although the position of the critical point can change. The associated strong singularity of the dynamics in the absence of noise leads to comparatively strong fluctuation effects.

Near the critical point, for small $Q,P$ the coordinate $Q$ varies in time much slower than $P$. This is seen from the linearized equations (\ref{eq:q_Langevin}), which in the absence of noise take the form $\dot Q= (\fp -1)Q -\mup P \sgn \gamma, \dot P=-(\fp+1)P +\mup Q\sgn\gamma$. 

Because $Q(\tau)$ is slow, function $P(\tau)$ adiabatically follows $Q(\tau)$. For the dimensionless time $\tau\gg 1$, function $P(\tau)$ can be expressed in terms of $Q(\tau)$ by setting $\dot P=0$. Substituting the result into the full nonlinear equation for $\dot Q$, one obtains
\begin{align}
\label{eq:slow_Q}
&\dot Q\approx -\partial_QU(Q) + \hat f_Q(\tau), \\
&U(Q)= \frac{1}{4}Q^2\left[\mup^2 - (\fp^2-1)\right]+\frac{1}{4}Q^4\left(-\mup + \frac{1}{3}Q^2\right)\nonumber
\end{align}
[$U(Q)$ has an extra factor $2/(\fp +1)$ which is set equal to 1, to the leading order in $\fp -1$]. We note that, because the relaxation time of $P(\tau)$ is small compared to that of $Q(\tau)$, fluctuations of $P(\tau)$ due to the quantum noise $\hat f_P(\tau)$, are small compared to fluctuations of $Q(\tau)$. This is the squeezing effect.

For $\mu_{B1}<\mup < \mu_{B2}$, the effective potential $U(Q)$ has a local maximum at $Q=0$ and two symmetric minima at 
$Q_{1,2}=\pm (\mu_{B}  + \mup)^{1/2}$.
The minima correspond to the stable period-two vibrations in the lab frame.  

For $\mup > \mu_{B2}, \fp^2>1$,  function $U(Q)$ still have the minima at $Q_{1,2}$, but now it has a local minimum at $Q=0$ and two additional local maxima at $Q=\pm (\mup -\mu_{B})^{1/2}$. The minimum at $Q=0$ corresponds to the stable zero-amplitude state. The local maxima correspond to unstable period-two vibrations.  

The behavior of the oscillator near the onset of period-two vibrations at $\mu_{B1}$ has similarities with the mean-field picture of the critical behavior at the second-order phase transition. The time-translation symmetry of the stable stationary state at $Q=0$ is spontaneously broken and there emerge two vibrational states with equal amplitudes and opposite phases and with period $4\pi/\omega_F$. Such behavior is usually described by an effective potential which is quartic in the coordinate $Q$.

A key observation is that, near the critical point $\fp=1,\mup=0$, it is necessary to keep the sixth-order term in $U(Q)$. This follows from Eq.~(\ref{eq:slow_Q}). The system becomes softer than at the bifurcation point far away from the critical point. Such form of the potential reminds the Landau free energy for a system that can undergo a first-order phase transition, and indeed, an analog of this transition can occur in the driven oscillator, see below. 

\subsection{The quantum Fokker-Planck equation}

Equation (\ref{eq:slow_Q}) has the form of the Langevin equation of a classical particle with one dynamical variable (half-degree of freedom) driven by a $\delta$-correlated noise. Indeed,  $\hat f_Q(\tau)$ and $Q(\tau)$ commute, and therefore $[Q(\tau),Q(\tau')]=0$.  However, the noise has quantum origin, its correlator (\ref{eq:noise_correlators}) explicitly contains $\hbar$. The slow quantum fluctuations can be alternatively described by the Fokker-Planck equation for the probability density $\rho(Q,\tau)$. From Eq.~(\ref{eq:slow_Q}), it reads
\begin{equation}
\label{eq:FPE}
\partial_\tau \rho = \partial_Q(\rho\,\partial_Q U) + {\cal D}\partial^2_Q \rho.
\end{equation}

\begin{figure}[h]
\includegraphics[width=7cm]{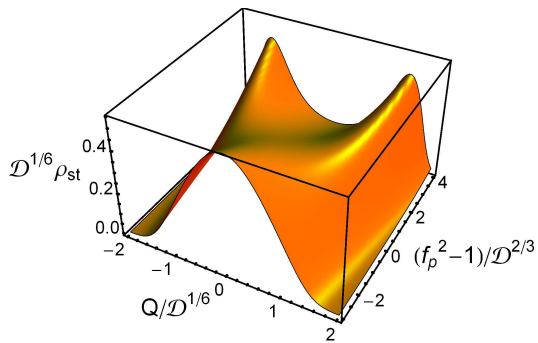}
\caption{Evolution of the stationary distribution $\rho_{\rm st}(Q)$ with varying amplitude of the driving field for $\mup=0$. For $\fp^2 < 1$ the distribution has a single broad maximum at $Q=0$. With increasing $\fp$ there are developed two sharp maxima at the positions that correspond to the period-two vibrational states.  }
\label{fig:contour}
\end{figure}

The stationary distribution of the oscillator near the critical point has the Boltzmann form
\begin{equation}
\label{eq:stationary_distribution}
\rho_{\rm st}(Q)=Z^{-1}\exp[-U(Q)/{\cal D}],
\end{equation}
where $Z=\int dQ\exp[-U(Q)/{\cal D}]$ is the effective partition function. The evolution of this distribution with varying $\fp$ is shown in Fig.~\ref{fig:contour}. 

The full stationary Wigner distribution is a product of $\rho_{\rm st}(Q)$ and a Gaussian distribution over $P$, with the position of the maximum over $P$ being $Q$-dependent.  Function $\rho_{\rm st}(Q)$ can be thought of as the integral over $P$ of this full distribution. The distribution (\ref{eq:stationary_distribution}) is maximal at the minima of $U(Q)$, i.e., at the values of $Q$ that correspond either to the stable states of the oscillator. Near the critical point the distribution is broad, which corresponds to large fluctuations in this range.

\section{Scaling of the rates of quantum-fluctuation induced switching}

Once the depth of the minima of $U(Q)$ becomes large compared to $\lap (2\bar n+1)$, the dynamics of the oscillator is characterized by two very different time scales. One is the relaxation time  $1/U''(Q)$ for $Q$ near the stable states. The other is a much slower time over which the oscillator can switch between the minima due to comparatively rare large fluctuations. The switching rate $W_{\rm sw}$  is given by the Kramers theory \cite{Kramers1940} of thermally activated switching over a potential barrier. This theory immediately extends in the present case to quantum fluctuations, even though there is, of course, no thermal activation for low temperatures.  In dimensionless time $\tau = \Gamma t$ 
\begin{align}
\label{eq:switching_rate}
&W_{\rm sw}=\Omega_{\rm sw}\exp(-R_A/\lap), \qquad R_A=\Delta U/(\bar n+1/2),\nonumber\\
&\Omega_{\rm sw}=[|U''(Q_{\cal S})|U''(Q_{\rm a})]^{1/2}/2\pi, \nonumber\\
&\Delta U=U(Q_{\cal S}) - U(Q_{\rm a}).
\end{align}
Here, $Q_{\rm a}$ is the position of the stable state (attractor) from which the oscillator switches and $Q_{\cal S}$ is the position of the adjacent local potential maximum [a saddle point on the $(Q,P)$-plane] over which the switching occurs.  Function $R_A$ is the quantum activation energy for the switching, it is proportional to the height of the potential barrier overcome in the switching.

Remarkably, Eq.~(\ref{eq:switching_rate}) for the switching rate has the same structure as the expression for the rate of tunneling from a potential well of $U(Q)$ in the sense that the rate is exponential in $1/\lap \propto 1/\hbar$. However, switching occurs not via tunneling under the barrier of $U(Q)$ but by going over the barrier. Tunneling requires the fast variable $P$ to be involved.  Therefore the tunneling exponent is parametrically larger than $R_A$ and the tunneling rate is exponentially smaller \cite{Marthaler2007a}.

The states between which the switching occurs and the rate $W_{\rm sw}$ depend on the parameter region on the bifurcation diagram in Fig.~\ref{fig:bifurcation}. We start with the region $\mu_{B1}(\fp) < \mup <\mu_{B2}(\fp)$. Here, the oscillator is bistable. It can switch from one period-two state to the other. By symmetry, the switching rates for the both states are the same. In Eq.~(\ref{eq:switching_rate}) $Q_{\rm a}= \pm (\mu_B +\mup)^{1/2}$, depending on the initially occupied state. The switching occurs over the unstable state $Q_{\cal S}=0$.

\begin{figure}[h]
\includegraphics[width=6cm]{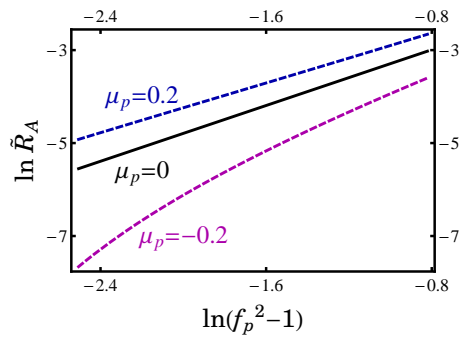}
\caption{The quantum activation energy $\tilde R_A=(\bar n + 1/2)R_A$ for switching between the parametrically excited period-two vibrational states with opposite phases. In the shown range  of the scaled driving field amplitude $1.04<\fp \equiv F/F_c < 1.2$ and the scaled frequency detuning $|\mup| <\mu_{B}(\fp)$, the period-two vibrations are the only stable states of the oscillator.  }
\label{fig:scaling}
\end{figure}

The dependence of the activation energy $R_A$ for switching between the period-two states on the parameters is shown in Fig.~\ref{fig:scaling}. For exact resonance, $\mup=0$, it follows from Eq.~(\ref{eq:slow_Q}) that $R_A= (\fp^2-1)^{3/2}/3(2\bar n +1)$. In this case $R_A$ displays a power law dependence on the distance from the critical field amplitude $f_p-1$. The scaling exponent is 3/2. For nonzero $\mup$, the behavior becomes more complicated and is not described by a simple power law. Close to the bifurcation line $\mu_{B1}(\fp)$, function $R_A$ scales as a power of the distance from  $\mu_{B1}(\fp)$,    $R_A \approx \mu_{B}(\mup-\mu_{B1})^2/2(2\bar n+1)$, cf. Ref.~\onlinecite{Dykman2007}.  We note that the scaling of the switching rates near a bifurcation point of a different type has been observed for resonantly driven Josephson junction based quantum oscillators \cite{Vijay2009}.

A different situation occurs in the range $\mup> \mu_{B2}(\fp)$ and $\fp >1$. Here, far from the bifurcation lines in Fig.~\ref{fig:bifurcation} the potential $U(Q)$ has  three local minima separated by two local maxima.
Interstate switching occurs via a transition from a minimum of $U(Q)$ over the adjacent maximum. This means that from the stable period-two states the oscillator can switch to the zero-amplitude state and vice versa, but direct switching between the period-two states is exponentially unlikely.

\begin{figure}[h]
\includegraphics[width=6cm]{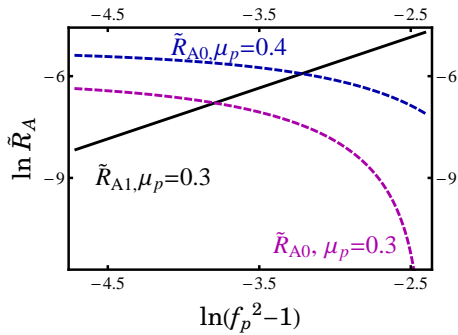}
\caption{The quantum activation energies $\tilde R_{A1}= (\bar n+1/2)R_{A1}$ and  $\tilde R_{A0}= (\bar n+1/2)R_{A0}$ 
for switching from a state of period-two vibrations to the zero-amplitude state and from the zero-amplitude state to one of the states of period-two vibrations, respectively. The curves refer to the region of tristability, where the scaled driving field amplitude $\fp  > 1$ and the scaled frequency detuning $\mup > \mu_{B}(\fp)$.  }
\label{fig:scaling_3states}
\end{figure}

The switching activation energies in the region of tristability are shown in Fig.~\ref{fig:scaling_3states};  $R_{A1}$ and $R_{A0}$ refer to switching from a period-two state to the zero-amplitude state and from the zero-amplitude state to one of the period-two states, respectively. To find $R_{A1}$, one sets in Eq.~(\ref{eq:switching_rate}) $Q_{\rm a} = \pm  (\mup + \mu_B)^{1/2}$, whereas to find $R_{A0}$ one sets $Q_{\rm a}=0$; in the both cases,  $Q_{\cal S}=\pm  (\mup - \mu_B)^{1/2}$. Interestingly, from Eq.~(\ref{eq:slow_Q}), the activation energy $R_{A1}$ displays simple scaling behavior, $R_{A1}= (\fp^2-1)^{3/2}/3(\bar n +1/2)$ independent of $\mup$. The activation energy of switching from the zero-amplitude state displays a more complicated behavior. It decreases with the increasing field amplitude. Near the bifurcation line $\mup=\mu_B(\fp)$ we have $R_{A0}\approx \mu_B(\mup-\mu_{B2})^2 /2(2\bar n+1)$, in agreement with the previous work \cite{Dykman2007}.

In the stationary regime, for $R_{A1}>R_{A0}$ it is exponentially more probable for the system to be in the state of period-two vibrations. On the other hand, for $R_{A0}>R_{A1}$, preferentially occupied is the zero-amplitude state, where the period-two vibrations are not excited. The interrelation between the field amplitude and frequency where $R_{A1}=R_{A0}$ corresponds to an analog of a first-order phase transition: here the populations of the different stable states of the oscillator are close to each other. 
From Eq.~(\ref{eq:slow_Q}), near the critical point the ``phase-transition" value of $\fp$ is given by expression $(\fp^2-1)^{1/2} = \mup/2$. It is shown by the dot-dashed line in Fig.~\ref{fig:bifurcation}.

\section{Quantum fluctuations in the immediate vicinity of the critical point}

Close to the critical point, where $|\fp^2-1|\lesssim {\cal D}^{2/3}, |\mup|\lesssim {\cal D}^{1/3}$, the stationary distribution (\ref{eq:stationary_distribution}) is almost flat, as seen from Fig.~\ref{fig:contour}. This indicates anomalously large quantum fluctuations of the oscillator variable $Q$ (whereas the fluctuations of the other variable, $P$, are squeezed). The fluctuational mean-square displacement is $\langle Q^2\rangle \sim {\cal D}^{1/3}$, whereas far from the critical region the mean-square displacement about a stable state is $\sim {\cal D}$ (${\cal D}\propto \hbar$ is the small parameter of the theory, ${\cal D}\ll 1$). In the critical region the exponents $R_A/\lap$ are no longer large and the concept of switching rates becomes ill-defined. One cannot separate switching from other fluctuations.  

A convenient characteristic of the fluctuation dynamics is the long-time decay of the correlation functions of the oscillator. This decay is characterized by the lowest nonzero eigenvalue $\nu_1$ of the Fokker-Planck equation (\ref{eq:FPE}). It can be found in a standard way \cite{vanKampen_book} by reducing Eq.~(\ref{eq:FPE}) to a Schr\"odinger-type equation for $\tilde\rho (Q) = \exp[U(Q)/2{\cal D}]\rho(Q)$. The quantum noise intensity ${\cal D}$ can be scaled out of this equation by changing to the scaled time ${\cal D}^{2/3}\tau $, scaled coordinate ${\cal D}^{-1/6}Q$, and scaled parameters ${\cal D}^{-1/3} \mup$ and ${\cal D}^{-2/3}(\fp^2 -1)$. 

\begin{figure}[h]
\includegraphics[width=6cm]{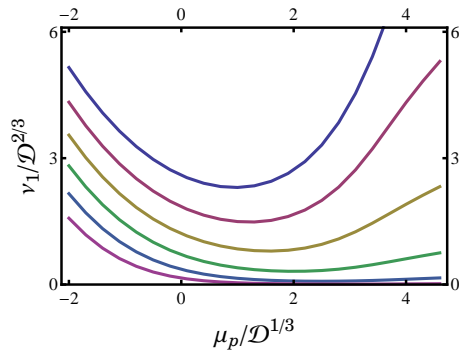}
\caption{The smallest nonzero eigenvalue $\nu_1$ of the kinetic equation (\ref{eq:FPE}), which determines the long-time decay of the correlation functions very close to the critical point $\mup=0,\fp=1$. The curves from top to down correspond to $(\fp^2-1)/{\cal D}^{2/3} = -4, -2, 0, 2, 4, 6$.  }
\label{fig:critical_decay}
\end{figure}

The dependence of the decrement $\nu_1$ on the frequency detuning parameter $\mup\propto (\omega_F-2\omega_0)/F_c$ for several values of the scaled driving field amplitude $\fp=F/F_c$ is shown in Fig.~\ref{fig:critical_decay}. It is seen that, for a given $\mup$, the dynamics is slowed down, i.e., $\nu_1$ decreases with the increasing driving amplitude. This is to be expected and is in agreement with Fig.~\ref{fig:contour}. For $\fp^2<1$ or $(\fp^2-1)^{1/2}+\mup <0$, the oscillator has one stable state, $Q=0$. As $\fp^2$ increases, the minimum of $U(Q)$ at $Q=0$ becomes more and more shallow and the relaxation rate, which is related to the curvature of $U(Q)$, goes down. As $\fp^2$ further increases, the system goes into the regime of coexisting period-two states (via the range of coexisting three stable states, for $\mup>0$, cf. Fig.~\ref{fig:bifurcation}). For large $(\fp^2-1)/{\cal D}^{2/3}$, $\nu_1$ is determined by the exponentially small rate of interstate switching that rapidly decreases with increasing $\fp$, see Figs.~\ref{fig:scaling} and \ref{fig:scaling_3states}.

The dependence of $\nu_1$ on the frequency detuning in Fig.~\ref{fig:critical_decay} is profoundly nonmonotonic. For $\fp^2-1<0$, where the system has only one stable state, $Q=0$, this behavior is a consequence of the nonmonotonic dependence of the curvature of $U''(0)$ on $\mup$, see Eq.~(\ref{eq:slow_Q}). Close to the critical point the nonparabolicity of $U(Q)$ is substantial. This is also seen from Fig.~\ref{fig:contour}, which shows that the stationary probability distribution becomes profoundly non-Gaussian near the critical point.  Therefore $\nu_1$ is different from $U''(0)$. The minimum of $\nu_1$ is reached for $\mup >0$. 

In the range $\mup>\mu_{B1}= -(\fp^2-1)^{1/2}$ the oscillator has two or three stable states in the neglect of fluctuations. For large $(\mup-\mu_{B1})/{\cal D}^{1/3}$, the decrement $\nu_1$ is determined by the rate of interstate switching.   This explains the  decrease of $\nu_1$ with increasing $\mup$ in Fig.~\ref{fig:critical_decay} for $\fp^2>1$ and for not too large $\mup/{\cal D}^{1/3}$.  

To understand the behavior of $\nu_1$ for larger $\mup$ in the range of tristability of the oscillator, we consider the balance equations for the populations of the two period-two states $w_{1,2}$ and the zero-amplitude state $w_0$, 
\begin{align}
\label{eq:balance}
&\dot w_0 = -2W_{01}w_0 + W_{10}(w_1+w_2), \nonumber\\
&\dot w_i=W_{01}w_0-W_{10}w_i \quad (i=1,2),
\end{align}
where $W_{01}=W_{02}\propto \exp(-R_{A0}/\lap)$ is the rate of switching from the zero-amplitude state to one of the period-two states and $W_{10}=W_{20}\propto \exp(-R_{A1}/\lap)$ is the rate of switching from a period-two state to the zero-amplitude state. 

From Eq.~(\ref{eq:balance}), $\nu_1=W_{10}$ in the region of tristability where the slow dynamics is determined by inter-state switching. The exponent $R_{A1}/\lap$ in Eq.~(\ref{eq:switching_rate})  for the rate $W_{10}$ is independent of $\mup$, cf. Fig.~\ref{fig:scaling_3states}. However, the prefactor  $\Omega_{\rm sw}\propto (\fp^2-1)^{1/2}[\mup^2-(\fp^2-1)]^{1/2}$ is increasing with increasing $\mup$. Therefore, somewhat unexpectedly, $\nu_1$ starts slowly increasing with $\mup$ in the tristability region.

\section{Conclusions}

The results of this paper refer to quantum fluctuations near the threshold of parametric excitation of a nonlinear oscillator. The threshold corresponds to the frequency of the driving field $\omega_F$ equal to twice the oscillator eigenfrequency $\omega_0$ and to the field amplitude $F$ equal to the critical value $F_c$ proportional to the oscillator decay rate.  For small $|F-F_c|, |\omega_F-2\omega_0|$, the oscillator dynamics is mapped onto the dynamics of an overdamped Brownian particle driven by quantum noise and confined in a symmetric potential well of the sixth order in the particle displacement.

In the immediate vicinity of the threshold quantum fluctuations are anomalously large and slow. The root-mean-square fluctuations of the amplitude scale with $\hbar$ and with temperature as $[\hbar (2\bar n +1)]^{1/6}$. The correlation time of the fluctuations scales as $[\hbar (2\bar n +1)]^{-2/3}$. We have found the dependence of the correlation time on the frequency detuning $\omega_F-2\omega_0$ and the field amplitude. Unexpectedly, the dependence on the frequency detuning turned out to be nonmonotonic.

For $F>F_c$, in the appropriate frequency range the oscillator can have coexisting states of period-two vibrations with the interstate distance in phase space significantly exceeding the mean-square-root fluctuations about the states, which are $\sim [\hbar (2\bar n +1)]^{1/2}$. Here, rare large quantum fluctuations lead to switching between the states. The switching mechanism is quantum activation. In switching the oscillator goes over an effective quasienergy barrier rather than tunneling beneath it. However, the logarithm of the switching rate is $\propto 1/\hbar$ for low temperatures, as in the case of tunneling. 

We found simple explicit expressions for the switching rates both in the regime where the period-two states are the only stable states of the oscillator and where the state with no period-two vibrations is also stable. In the region of bistability, at exact resonance the switching activation energy scales with the field amplitude as $(F-F_c)^{3/2}$. This scaling applies also to the switching from period-two states in the regime of tristability for arbitrary detuning. In contrast, the dependence of the activation energy on $F-F_c$ in the regime of bistability away from exact resonance, and the dependence on $F-F_c$ of the activation energy of switching from the zero-amplitude state do not display simple scaling, generally. In the region of tristability, depending on the field parameters, either the period-two states or the zero-amplitude state are predominantly occupied. The state occupations are close in a narrow parameter range, the behavior that reminds the first-order phase transition.

\acknowledgements
ZRL was supported by the RIKEN FPR program. YN was supported in part by ImPACT Program of Council for Science, Technology and Innovation (Cabinet Office, Government of Japan), the Project for Developing Innovation System of MEXT, JSPS KAKENHI (Grant No. 26220601), and National Institute of Information and Communications
Technology (NICT). MID was supported in part by the U.S. Army Research Office (W911NF-12-1-0235), the U.S. Defense Advanced Research Agency (FA8650-13-1-7301), and the  NSF.


%

\end{document}